\documentclass[12pt]{article}
\usepackage{amsmath}
\usepackage{amssymb}
\usepackage{amsfonts}
\newcommand{\ba}{\begin{array}{l}}
\newcommand{\ea}{\end{array}}
\newcommand{\beq}{\begin{equation}}
\newcommand{\eeq}{\end{equation}}
\newcommand{\bea}{\begin{eqnarray}}
\newcommand{\eea}{\end{eqnarray}}
\usepackage{epsfig}
\usepackage{graphicx}
\usepackage{color}

\definecolor{dyellow}{rgb}{1.,0.8,.0}
\definecolor{myblue}{rgb}{.1,.1,.7}
\definecolor{dcyan}{rgb}{.0,.6,.6}
\definecolor{dmagenta}{rgb}{0.6,0.0,0.6}
\definecolor{brown}{rgb}{0.6,0.2,0.}
\definecolor{darkblue}{rgb}{.0,.0,0.5}
\definecolor{darkred}{rgb}{0.75,0.0,0.0}
\definecolor{orange}{rgb}{1.,.6,.0}
\definecolor{dorange}{rgb}{0.8,.4,.0}
\definecolor{darkgreen}{rgb}{0.0,0.6,0.0}
\definecolor{purple}{rgb}{.4,.0,.4}

\def\bc{\begin{center}}
\def\ec{\end{center}}
\def\be{\begin{eqnarray}}
\def\ee{\end{eqnarray}}

\newcommand{\omits}[1]{}

\begin{document}
\begin{center}
{\Large \bf { Kinematics in Randers-Finsler geometry \\and\\ secular increase of the astronomical unit}}\\
  \vspace*{1cm}
Xin Li$^{\ast,\dagger}$ \footnote{lixin@itp.ac.cn} and Zhe Chang$^{\ast,\ddagger}$ \footnote{changz@mail.ihep.ac.cn}\\
\vspace*{0.2cm} {\small $^\ast$Theoretical Physics Center for Science Facilities, Chinese Academy of Sciences}\\
{\small $^\dagger$Institute of Theoretical Physics,
Chinese Academy of Sciences, 100190 Beijing, China}\\
{\small $^\ddagger$Institute of High Energy Physics, Chinese Academy
of Sciences, 100049 Beijing, China}\\

\bigskip

\end{center}
\vspace*{2.5cm}

%
\begin{abstract}\baselineskip=30pt
Kinematics in Finsler space is investigated. It is showed that the result based on the kinematics with a special Finsler structure is in good agreement with the reported value of secular trend in the astronomical unit, $d{\rm AU}/dt=15\pm4[{\rm m/century}]$. The space deformation parameter $\lambda$ in this special structure is very small with scale of $10^{-6}$ and should be a constant. This fact is consistent with the reported value of an anomalous secular eccentricity variation of the Moon's orbit.
 \vspace{1cm}
\begin{flushleft}
Key words: Finsler geometry, kinematics, astronomical unit\\
PACS numbers:  02.40.-k, 04.50.Kd, 95.10.Eg
\end{flushleft}
\end{abstract}

\newpage
\baselineskip=30pt
The rapid progress in technology make the astronomical observations more and more accurate. New physical phenomena have appeared, which can not be explained by conventional physical mechanisms. The most well-known, among them, is the accelerated expanding universe \cite{Riess} and the flat rotational velocity curves of spiral galaxies \cite{Trimble}. The astronomical units (AU) is the fundamental and standard scale in astronomy. The latest planetary ephemerides \cite{Pitjeva} presented the accurate value of AU with tiny error
 \be\label{AU}
 1[{\rm AU}]=1.495978706960\times10^{11}\pm0.1[{\rm m}].
 \ee
However, the recent reports from Krasinsky and Brumberg \cite{Krasinsky} and also from Standish \cite{Standish} show a positive secular trend in AU as $d{\rm AU}/dt=15\pm4~[{\rm m/century}]$. These authors have analyzed all available radiometric measurements on distances between
the Earth and the inner planets including observations of martian landers and orbiters. This value is about 100 times larger than the current determination error of AU \cite{Pitjeva}. The theoretical value of round-trip time of radar signal is given as
 \be t_{\rm theo}=\frac{d_{\rm theo}[{\rm AU}]}{c}, \ee where $d_{\rm theo}$ is interplanetary distance obtained
from ephemerides and $c$ is the speed of light. The secular trend was obtained by the following formula \be t_{\rm theo}=\frac{d_{\rm theo}\left[{\rm AU}+\frac{d{\rm AU}}{dt}(t-t_0)\right]}{c},\ee where $t_0$ is the initial epoch. Currently, none of theoretical predictions is consistent with the time dependent term $\frac{d{\rm AU}}{dt}(t-t_0)$. To explain this fact, physicists have made several attempts, such as the effects of the cosmic expansion \cite{Krasinsky,Mashhoon,Arakida1}, the time variation of gravitational constant \cite{Krasinsky}, mass loss of the Sun \cite{Krasinsky,Noerdlinger}, and the influence of dark matter on light propagation in the solar system \cite{Arakida2}. However, none of them seems to be successful. Recently, one sound model have been proposed \cite{Miura}. It assumes the existence of some tidal interactions that transfer the angular momentum from the Sun to the planets system, and makes use of the conservation law of total angular momentum to explain the secular trend in AU. This model also need more work before it can be considered to be viable.

As we mentioned in the beginning of this Letter, general relativity faces problems indeed. Also, analyzing the data from pioneer 10 and 11 spacecraft show that an anomaly acceleration exists in the solar system, which can not be explained by the Newtonian gravity and general relativity \cite{Anderson}. Finsler geometry is a natural generation of Riemann geometry. The gravitational theory based on Finsler geometry supplies a reasonable way to solve the problems mentioned above. In a previous paper \cite{Finsler DE}, we proposed a modified Friedmann model based on the Einstein equations in Finsler space, which guarantees an accelerated expanding universe without invoking dark energy. Also, in the framework of Finsler geometry, the flat rotation curves of spiral galaxies can be deduced naturally without invoking dark matter \cite{Finsler DM}. Special relativity in Randers (a special kind of Finsler space) \cite{Randers} has been investigated \cite{RF}. We found that the anomalous acceleration observed by Pioneer 10 and 11 is corresponded with a special structure of Randers space \cite{Finsler PA}.

In this Letter, we will discuss the secular trend in AU in the frmework of Finsler geometry. We notice the difference between lengths in Finsler geometry and Riemann geometry. The length in
Riemann geometry is a function of positions. However, this is not the
case in Finsler geometry. In Finsler geometry, the length is a
function of both positions and velocities. Finsler geometry is base on
the so called Finsler structure $F$ with the following property
$F(x,\lambda y)=\lambda F(x,y)$, where $x\in M$ represents the position
and $y\in T_xM$ represent velocity, $M$ is an n-dimensional manifold. The Finslerian metric is given as \cite{Book
by Bao}
 \be
 g_{\mu\nu}\equiv\frac{\partial}{\partial
y^\mu}\frac{\partial}{\partial y^\nu}\left(\frac{1}{2}F^2\right).
\ee

Then, $g_{\mu\nu}$ defines a Riemannnian metric
\be g_{\mu\nu}dy^\mu\otimes dy^\nu \ee on the punctured tangent space $T_xM\backslash0$ \cite{Bao}. It admits the unit tangent sphere (or indicatrix) $I_xM\equiv\{y\in T_xM:F(y)=1\}$ as smooth Riemannian manifold. Topologically, $I_xM$ is diffeomorphic to unit sphere $S^{n-1}$ in $R^n$. The volume form of the indicatrix $I_xM$ is
\be\label{vol of F} \sqrt{g}\sum_{\mu=1}^n(-1)^{\mu-1}\frac{y^\mu}{F}dy^1\wedge\cdots\wedge dy^{\mu-1}\wedge dy^{\mu+1}\wedge\cdots\wedge dy^n, \ee where $g$ denotes the determinant of the metric $g_{\mu\nu}$.

All the trajectories of planets in the solar system almost lie in the same plane, the eccentricity of planets (exclude Mercury and Pluto) is very small. Thus, the trajectories of planets can be considered as circular orbit which embedded in three dimensional space. It is well-known that in Euclidean space the length of unit circle equals $2\pi_E$ or the value of $2\times3.1415926\cdots$. However, in Finsler space it is typically not equal to $2\times3.1415926\cdots$. In Finsler space, as mentioned in (\ref{vol of F}) the 2-dimensional indicatrix $I^2_xM$ has length element
\be ds=\frac{\sqrt{g}}{F}\left(y^1\frac{dy^2}{dt}-y^2\frac{dy^1}{dt}\right)dt, \ee where $t$ is a real parameter. Then, the length of the indicatrix $I^2_xM$ is
\be L\equiv\int_{F=1}ds.\ee Here, we confine the Finsler structure $F$ as Randers type
\be\label{Randers type} F=\sqrt{(y^1)^2+(y^2)^2}+\lambda y^1,\ee where the parameter $\lambda$, in general, is a function of positions. Introduce polar coordinates on $I^2_xM$, $y^1=r\cos\phi$ and $y^2=r\sin\phi$. The determinant $g$ in polar coordinate is
\be  g=\left(\frac{1}{r}\right)^3.\ee Thus, the length $L$ is of the form
\be\label{length} L=\int_0^{2\pi_E} \frac{1}{\sqrt{1+\lambda\cos\phi}}d\phi.\ee
Here, we can see from the integral (\ref{length}) that the length $L$ equals $2\pi_E$ only if $\lambda$ takes value of 0. Since the Modification on the Newton's gravity is very tiny, we suppose that $\lambda$ is very small. Hence, to second order in $\lambda$, the length can be derived as \be\label{pi of F} L&=&\int_0^{2\pi_E}\left(1-\frac{\lambda}{2}\cos\phi+\frac{3\lambda^2}{8}\cos^2\phi\right)\nonumber\\
     &=&2\pi_E\left(1+\frac{3\lambda^2}{16}\right).
\ee
The equation (\ref{pi of F}) tells us  that the value of $\pi$ in Finsler geometry is
\be \pi_F=\pi_E\left(1+\frac{3\lambda^2}{16}\right).\ee

In Newton's gravity, the orbital angular momentum of planet is conserved. In other word, the second law of Kepler is valid. The line joining a planet and the Sun sweeps out equal areas during equal intervals of time, $\frac{dA}{dt}=L/2m$. Where $A$ is the areas, $L$ is the orbital angular momentum and $m$ is the mass of the planet. However, the report from Krasinsky and Brumberg \cite{Krasinsky} implies that during equal intervals of time the area sweep out by the line joining a planet and the Sun is increasing. Unlike the explanation of Miura \cite{Miura}, we attribute this phenomenon to the different value of $\pi$ in Finsler geometry. In Finsler space of Randers type, the area of disk with boundary ($F=R$) is $\pi_FR^2$. Then, the difference between the areas of disk in Riemann geometry and Finsler geometry is
\be\label{delta A1} \delta{A}\equiv\pi_FR^2-\pi_ER^2=\frac{3\lambda^2}{16}\pi_ER^2. \ee Making use of the result of Krasinsky and Brumberg, we get
\be \frac{d^2A}{dt^2}=\frac{\dot{L}}{2m}=\frac{\sqrt{GM_\odot/R}}{4}\dot{R},\ee where the dot denotes derivative respect to time and $M_\odot$ is the mass of the Sun. Hence, the increased area of disk in one orbital periods of planet is
\be\label{delta A2} \delta{A}=\frac{1}{2}\frac{d^2A}{dt^2}T^2=\left(\frac{\pi_E\dot{R}}{2\sqrt{GM_\odot/R}}\right)\pi_ER^2,\ee where $T$ is the orbital periods of planet and we used the Kepler's third law to get the second equation of (\ref{delta A2}). Combining the equation (\ref{delta A1}) and (\ref{delta A2}), we obtain
\be\label{lambda} \lambda=\sqrt{\frac{4\dot{R}}{3R}T}. \ee
Here, by taking the average value of $d{\rm AU}/dt$, we list the values of the $\lambda$ for each planet of the solar system respectively in Table 1.

The values of the $\lambda$ given in Table 1 are very close for inner planets. One should notice that the analyzed data of distances \cite{Krasinsky} is in the range of inner planets and Martian landers and orbiters. This fact implies that the space deformation parameter $\lambda$ should be a constant in the solar system. By analyzing the data of Table 1, we obtain the value of the constant parameter $\lambda=1.0776\times10^{-6}$.

A recent orbital analysis of Lunar Laser Ranging (LLR) \cite{Williams} shows an anomalous secular eccentricity variation of the Moon's orbit $\rm (0.9\pm0.3)\times10^{-11}/yr$, equivalent to an extra $\rm 3.5~mm/yr$ in perigee and apogee distance \cite{Anderson1}. By supposing the variation of the distance from the center to focus and the semi-major axis be the same in the moon orbit, namely $\delta a=\delta c$, we obtain
\be\label{delta e} \delta a= \frac{a\delta e}{1-e}. \ee Here, $a$ denotes the semi-major axis and $e$ denotes eccentricity. By making use of equation (\ref{delta e}) and the observation data of LLR, we obtain the secular variation of Moon orbital semi-major axis as
 \be\label{result delta a} \delta a=3.62\pm1.20 {\rm~ mm/yr}. \ee

Under the premise that $\lambda$ is constant, and by making use of equation (\ref{lambda}), we obtain the secular variation of Moon orbital radius $\dot{R}_{M}=4.48~{\rm mm/yr}$. This result is consistent with formula (\ref{result delta a}). Thus, our hypothesis that the parameter $\lambda$ is constant is supported by the observation of LLR.

The uniform space deformation means that the secular trend of planetary orbits are $\dot{R}_{Pl}\propto R^{-1/2}$. We list the values of $\dot{R}_{Pl}$ for each planet of the solar system respectively in Table 2. We wish this could be tested in future astronomical observations.

\begin{table}
\caption{The values of semi-major axis of the planetary orbit $a_{PL}$ and orbital periods of planets $T_{PL}$ given in Ref.\cite{Pitjeva}. The space deformation parameter $\lambda$ refers to different planets are listed. }
\begin{center}
\begin{tabular}{c|ccc}
\hline
Planets & $a_{PL}$(AU) & $T_{PL}$(years) & $\lambda$($10^{-6}$)\\ \hline
Mercury & 0.38709893 & 0.240840253   &0.910799787\\
Venus   & 0.72333199 & 0.615171854   &1.064875317\\
Earth   & 1.00000011 & 1.0           &1.154700475\\
Mars    & 1.52366231 & 1.880815968   &1.282915811\\
Jupiter & 5.20336301 & 11.85631638   &1.743019298\\
Saturn  & 9.53707032 & 29.42300556   &2.028174849\\
Uranus  & 19.19126393& 84.01058299   &2.415931289\\
Neptune & 30.06896348& 164.7856303   &2.703147899\\
Pluto   & 39.48168677& 247.6753567   &2.892097417\\ \hline
\end{tabular}
\end{center}
\end{table}

\begin{table}
\renewcommand{\arraystretch}{1.2}
\caption{In the case of the uniform space deformation ($\lambda=1.0776\times10^{-6}$), the secular trend of each planet orbit are listed. }
\begin{center}
\begin{tabular}{c|c}
\hline
Planets & $\dot{R}_{PL}$[m/century] \\ \hline
Mercury & 21.0 \\
Venus   & 15.4 \\
Earth   & 13.1 \\
Mars    & 10.6 \\
Jupiter & 5.73 \\
Saturn  & 4.23 \\
Uranus  & 2.98\\
Neptune & 2.38\\
Pluto   & 2.08\\ \hline
\end{tabular}
\end{center}
\end{table}
\bigskip

\centerline{\large\bf Acknowledgements} \vspace{0.5cm}
 We would like to thank Prof. C. J. Zhu, H. Y. Guo and C. G. Huang for useful discussions. The
work was supported by the NSF of China under Grant No. 10525522 and
10875129.


\begin{thebibliography}{999}
\bibitem{Riess}A. G. Riess, {\it et al}.,
Astrophys J. {\bf 117} (1999) 707; S. Perlmutter, {\it et al}.,
Astrophys J. {\bf 517} (1999) 565; C. L. Bennett, {\it et al}.,
Astrophys J. {\bf 148} (Suppl.) (2003) 1.
\bibitem{Trimble}V. T. Trimble, Annu. Rev. Astron. Astrophys. {\bf 25} (1987) 425.
\bibitem{Pitjeva}E. V. Pitjeva, Solar System Research {\bf 39} (2005) 176.
\bibitem{Krasinsky}G. A. Krasinsky and V. A. Brumberg, Celest. Mech. Dyn. Astrn. {\bf 90} (2004) 267.
\bibitem{Standish}E. M. Standish, Proc. IAU Colloq. {\bf 196} (2005) 163.
\bibitem{Mashhoon}B. Mashhoon, N. Mobed and D. Singh, Class. Quant. Grav. {\bf 24} (2007) 5031.
\bibitem{Arakida1}H. Arakida, New Astron. {\bf 14} (2009) 275.
\bibitem{Noerdlinger}P. D. Noerdlinger, arXiv:astro-ph/0801.3807.
\bibitem{Arakida2}H. Arakida, arXiv:astro-ph/0810.2827.
\bibitem{Miura}T. Miura {\it et al}., arXiv:astro-ph.EP/0905.3008.
\bibitem{Anderson}J. D. Anderson, {\it et al}., Phys. Rev.
Lett. {\bf 81} (1998) 2858, J. D. Anderson, {\it et al}.,
Phys. Rev. D {\bf 65} (2002) 082004, J. D. Anderson, {\it et al}.,
Mod. Phys. Lett. A {\bf 17} (2002) 875.
\bibitem{Finsler DE}Z. Chang and X. Li, Phys. Lett. B {\bf 676} (2009) 173.
\bibitem{Finsler DM}Z. Chang and X. Li, Phys. Lett. B {\bf 668} (2008) 453.
\bibitem{Randers}G. Randers, Phys. Rev. {\bf 59} (1941) 195.
\bibitem{RF}Z. Chang and X. Li, Phys. Lett. B {\bf 663} (2008) 103.
\bibitem{Finsler PA}X. Li and Z. Chang, arXiv:astro-ph/0909.3713.
\bibitem{Book by Bao}D. Bao, S. S. Chern and Z. Shen, {\it An
Introduction to Riemann--Finsler Geometry}, Graduate Texts in
Mathmatics {\bf 200}, Springer, New York, 2000.
\bibitem{Bao}D. Bao and Z. Shen, Results in Math. {\bf 26} (1994) 1.
\bibitem{Williams}J. G. Williams and D. H. Boggs, in Proceedings of 16th International Workshop on Laser
Ranging ed. S. Schillak, (Space Research Centre, Polish Academy of Sciences), 2009.
\bibitem{Anderson1}J. D. Anderson and M. M. Nieto, arXiv:gr-qc/0907.2469.








\end{thebibliography}
\end{document}